\documentclass[11pt]{article}
\usepackage{jheppub}
\usepackage[dvipsnames]{xcolor}
\usepackage{amsfonts,amsmath,amssymb,bm}
\usepackage{comment}
\usepackage{graphicx}
\usepackage{multirow}
\usepackage{booktabs}

\makeatletter
\def\@fpheader{\relax}
\makeatother

\title{Machine learning automorphic forms for black holes}

\author[*]{Vishnu Jejjala}
\author[\dagger]{\!\!, Suresh Nampuri}
\author[*]{\!\!, Dumisani Nxumalo}
\author[*]{\!\!, Pratik Roy}
\author[*]{\!\!, Abinash Swain}

\affiliation[\, *]{Mandelstam Institute for Theoretical Physics, School of Physics,  and NITheCS,\\ University of the Witwatersrand, Johannesburg, WITS 2050, South Africa}
\affiliation[\, \dagger]{Center for Mathematical Studies, University of Lisbon (CEMS.UL), UID/04561/2025 FCT, Portugal}

\emailAdd{vishnu.jejjala@gmail.com}
\emailAdd{nampuri@gmail.com}
\emailAdd{9510671P@students.wits.ac.za}
\emailAdd{pratik.roy@wits.ac.za}
\emailAdd{2642931@students.wits.ac.za}

\newcommand\capt[1]{\caption{\textsf{#1}}}
\newcommand\be{\begin{equation}}
\newcommand\ee{\end{equation}}
\newcommand\bea{\begin{eqnarray}}
\newcommand\eea{\end{eqnarray}}

\newcommand\eref[1]{\eqref{#1}}

\newcommand\ra[1]{\renewcommand{\arraystretch}{#1}}
\newcommand\abs[1]{|#1|}

\abstract{
Modular, Jacobi, and mock-modular forms serve as generating functions for BPS black hole degeneracies. By training feed-forward neural networks on Fourier coefficients of automorphic forms derived from the Dedekind eta function, Eisenstein series, and Jacobi theta functions, we demonstrate that machine learning techniques can accurately predict modular weights from truncated expansions. Our results reveal strong performance for negative weight modular and quasi-modular forms, particularly those arising in exact black hole counting formul\ae, with lower accuracy for positive weights and more complicated combinations of Jacobi theta functions. This study establishes a proof of concept for using machine learning to identify how data is organized in terms of modular symmetries in gravitational systems and suggests a pathway toward automated detection and verification of symmetries in quantum gravity.
}

\begin{document}

\maketitle

\section{Introduction}
Over the last two decades, automorphic forms have emerged as encoders of the mathematical principles underlying the organization of information and microstates in superstring approaches to quantum gravity.
Perhaps their most significant appearance in this context\footnote{See~\cite{Sen:2007qy,Murthy:2023mbc,Alexandrov:2025sig} for reviews.} is the case of four-dimensional \cite{Dabholkar:2004yr,Dabholkar:2005by,Jatkar:2005bh,Dabholkar:2012nd, Dabholkar:2014ema} and five-dimensional~\cite{Maldacena:1999bp} supersymmetric (BPS) black hole microstate counting. The black holes are extremal solutions of the low-energy gravitational theories in four-dimensional and five-dimensional superstring or M-theory compactification models and consequently support an AdS$_2$ or AdS$_3$ factor in their near-horizon geometry~\cite{Kunduri:2007vf}, which itself is a full solution of the gravitational equations of motion and encodes all features pertaining to the black hole horihorizon,luding the black hole entropy, independent of the asymptotic boundary conditions.
Consequently, all field content determining the black hole entropy in the near-horizon geometry can be shown to depend only on the charge parameters describing the black hole~\cite{Ferrara:1996dd,Ferrara:1996um,Sen:2005wa,Goldstein:2005hq,Dabholkar:2006tb}. The corresponding quantum statistical counting function that generates the integral microstate degeneracies in terms of quantized integral charges must therefore transform covariantly under the discrete version of the isometries of the near-horizon region.

As these isometries contain an $SL(2,\mathbb{Z})$ factor, we expect the generating functions to be automorphic forms (modular, mock-modular, or Jacobi forms) of this group or its principal congruence subgroups, depending on the theory under consideration. This expectation has been spectacularly vindicated by the exact counting generating functions for classes of BPS black holes, the $\frac18$-BPS solutions in $\mathcal{N}=8$ supergravity~\cite{Maldacena:1999bp} and both $\frac12$-BPS~\cite{Dabholkar:2004yr} and $\frac14$-BPS solutions~\cite{Dijkgraaf:1996it,Maldacena:1999bp,Jatkar:2005bh} in  $\mathcal{N}=4$ theories in four dimensions, identified as Jacobi, modular, and Siegel modular forms, respectively.

To be precise, upon compactifying type II superstring theory on K3$\,\times\, T^2$ or heterotic string theory on $T^6$, we enumerate the degeneracies of BPS solutions by calculating the Fourier coefficients of modular, Jacobi or Siegel modular forms. For instance, in the case of $\mathcal{N}=4$ supersymmetry, the degeneracies of $\frac14$-BPS states are encoded by the inverse of the Igusa cusp form $\Phi_{10}(\tau,z,\sigma)$, a Siegel modular form. Its Fourier coefficients in the $z$-expansion yield Jacobi forms whose coefficients count BPS states with fixed charges. These modular symmetries arise from the duality group acting on the charge lattice. The degeneracies are extracted \cite{LopesCardoso:2004law,LopesCardoso:2021aem} via a contour integral~\cite{Dijkgraaf:1996it}:
\begin{equation}
d(Q_e, Q_m) \sim \int d\tau\, dz\, d\sigma\ e^{-\pi i (\tau Q_e^2 + \sigma Q_m^2 + (2z-1) Q_e\cdot Q_m)} \frac{1}{\Phi_{10}(\tau,z,\sigma)}~,
\end{equation}
where $Q_e$ and $Q_m$ are the electric and magnetic charge vectors, respectively. This can be compared to the Bekenstein--Hawking entropy, including one-loop corrections. Modular invariance helps organize the degeneracies, ensures duality invariance, and simplifies computations via the Rademacher expansion. The structure of the modular form controls the asymptotic growth of states:
\begin{equation}
\log d(Q_e, Q_m) \sim \pi \sqrt{Q_e^2 Q_m^2 - (Q_e\cdot Q_m)^2}~,
\end{equation}
matching the classical area law~\cite{Cvetic:1995uj}.

In certain cases, the generating functions of black hole microstates are not modular forms but mock-modular forms~\cite{Dabholkar:2012nd}. For single-centred black holes in $\mathcal{N}=4$ supergravity, the counting function is not fully modular because of the wall crossing phenomenon and contributions from multi-centred configurations.
Mock-modularity arises when polar terms associated with multi-centred states are subtracted to isolate the single-centred degeneracies.
These functions admit a modular completion from the inclusion of a ``shadow'' and are essential for understanding non-perturbative contributions to the black hole entropy. The entropy formula involves mock-modular growth and is still governed by the Hardy--Ramanujan--Rademacher expansion and its generalizations.

We recall that a modular form $\phi_k(\tau)$ of weight $k\in \mathbb{Q}$ is a function of a complex variable $\tau = \tau_1+ i \tau_2$, defined on the upper half plane ($\tau_2 > 0$), which transforms under an $SL(2,\mathbb{Z})$ operation as 
\begin{equation}
\phi_k\left(\frac{a\tau + b}{c\tau +d}\right) = (c \tau +d)^k \phi_k(\tau)~,\quad \bigg(\begin{array}{cc} a & b \\ c & d \end{array}\bigg) \in SL(2, \mathbb{Z})~.
\end{equation}
Jacobi forms are two variable generalizations of modular forms with an additional complex variable $z$ associated with an index $m$.
These transform as 
\begin{eqnarray}
    \phi_{k,m}\left(\frac{a\tau + b}{c\tau +d}, \frac{z}{c\tau +d}\right)& = &(c \tau +d)^k e^{2 \pi i \frac{m c z^2}{c \tau + d}}\phi_{k,m}(\tau,z)~,\\
    \phi_{k,m}(\tau, z + \lambda \tau + \mu)&=& e^{-2 \pi i m (\lambda^2 \tau + 2 \lambda z + \lambda \mu)} \phi_{k,m}(\tau, z)~, \quad \lambda, \mu \in \mathbb{Z}~,\nonumber
\end{eqnarray} 
while Siegel modular forms are a genus two version of modular forms defined on the Siegel upper half plane (UHP) parametrized by
\begin{equation}
\Omega = \bigg(\begin{array}{cc} \rho & v \\ v & \sigma \end{array}\bigg)~,
\end{equation}
which transforms with a weight $k$ under $Sp(2,\mathbb{Z})$: 
\begin{equation}
\Phi_k\left(\frac{a\Omega+b}{c \Omega + d}\right)= (c \,\Omega + d)^k \Phi_k(\Omega)~. 
\end{equation}
Notable examples of modular forms include the Dedekind $\eta(\tau)$ function of weight $\frac12$, the four Jacobi theta functions $\theta_{i}(\tau,z),\, i \in \{1,2,3,4\}$, and the Igusa cusp form of weight $10$ that we have mentioned above. Since these appear in various black hole counting formul\ae, these modular forms will be the focus of our work. 

In addition, there exists another modular quantity of relevance to black hole counting, namely the quasi-modular form. This does not possess full modularity, but nevertheless transforms in a well-defined manner (parametrized by a depth $s$) under modular transformations as 
\begin{equation}
\mu_k\left(\frac{a \tau +b}{c \tau + d}\right) = (c\tau+d)^k \left(\mu_k(\tau) + \sum^s_{i=1} \alpha_i(\tau) \left(\frac{c}{c\tau+d}\right)^i\right)~.
\end{equation}

The modular forms of interest in this paper have a known convergent Fourier expansion about $\tau = i \infty$:
\begin{eqnarray}
    \phi_k(\tau) = \sum_{n\geq n_0} a_n q^n,
\end{eqnarray}
with $q = e^{2 \pi i \tau}$.
If $n_0 > 0$, $\phi \rightarrow 0$ as $\tau \rightarrow i \infty$ and is referred to as a cusp form, whereas $n_0 < 0$ implies a pole at $\tau = i \infty$, and it is referred to as a weakly holomorphic modular form.
In the latter case, the asymptotic growth of its coefficients goes as $a_n \rightarrow e^{\sqrt{n}}$, and as black hole degeneracies are exponential in integer charges, weakly holomorphic forms emerge as counting functions of their degeneracies.  

Microscopic generating functions, as mentioned above, which have been exactly written down for a class of supersymmetric black holes in $\mathcal{N}=4$ and $\mathcal{N}=8$ superstring heterotic compactification models, are expressed in terms of negative weight weakly holomorphic modular forms such as $\frac{1}{\Delta}=\frac{1}{\eta^{24}(\tau)}$, quasi-modular forms as $\frac{E_2(\tau)}{\eta^{24}(\tau)}$ of depth $1$, and weakly holomorphic Jacobi forms derived from the inverse of Igusa cusp form as well as negative weight Jacobi forms such as $\frac{\theta_1^2(\tau,z)} {\eta^6(\tau)}$~\cite{Dabholkar:2012nd}. These have known Fourier expansions in the $\tau$ upper half plane around $\tau_2 = \infty$ or equivalently, around $q = e^{2\pi i \tau}= 0$. Crucially, when the expansion has a finite number of terms with negative powers of $q$, referred to as polar terms, each of the coefficients of the countably infinite remaining terms can be exactly expressed in terms of the polar coefficients by a Rademacher expansion. This implies that the degrees of freedom relevant to the organization of microstates in the classes of black holes amenable to counting by these automorphic forms are fully encoded in the polar part, and constitutes but one example of deep physical insights that the mathematical rules governing modular forms translate into. Other important insights from analyzing the modular properties include the identification of loci of marginal stability governing black hole decays as well as the contribution of wormholes to supersymmetric black hole counting. In sum, a subset of the rules of quantum gravity is encoded in the language of modular symmetries. 
Hence, detecting and identifying modular forms and analyzing their modular properties is a clear operational pathway to understanding the Hilbert space of quantum gravity. 

Machine learning techniques have proven to be adept at pattern recognition. Indeed, the role of neural networks is to identify subtle correlations among the columns in a spreadsheet. The problem of identifying modular symmetries reduces to recognizing patterns in number theory, the ``queen of mathematics'' in Gauss's parlance.
This is one of the areas of mathematics where artificial intelligence has to date made the least headway.\footnote{Among the few remarkable results in this area are in machine learning aspects of the Sato--Tate conjecture~\cite{He:2020kzg}, predicting invariants of arithmetic curves~\cite{He:2020tkg}, and establishing the existence of striking patterns of murmurations embedded within the statistics of large families of elliptic curves~\cite{He:2022pqn}.} This renders the search for modular symmetries relevant to quantum gravity both challenging and important. We take preliminary steps in this direction by developing neural network protocols for classifying various automorphic forms and report on the efficacy of these methods as well as the immediate potential applications in black hole counting.

The organization of this paper is as follows. In Section~\ref{sec:results}, we report that we are able to identify black hole counting functions from the Fourier coefficients in their expansions. Our focus is on modular, Jacobi, and quasi-modular forms of $SL(2,\mathbb{Z})$. In Section~\ref{sec:summary}, we explain the significance of these findings and point to future work. Appendix~\ref{app:af} collects definitions of automorphic forms.
Appendix~\ref{app:ml} describes the data structure and neural network architecture we employ in our experiments.

\section{Machine learning automorphic forms from Fourier coefficients}\label{sec:results}

In both gravitational and microscopic calculations of BPS black hole entropy, in theories with symmetries comprising $SL(2,\mathbb{Z})$ or one of its congruent subgroups, the counting function is expected to be an automorphic form (modular, Jacobi, Siegel, or mock-modular). However, we may be able to computationally access and write down only approximate counting functions that encode the leading entropy along with potentially a finite number of subleading terms in a large charge expansion. These would correspond to a finite amount of data regarding the putative automorphic form, such as a finite number of terms in the Fourier expansion of a modular form around $\tau = i \infty$. Hence, given the \textit{a priori} knowledge of a counting function being, say, a modular form, a finite number of whose terms can be computed, it is expedient to examine whether we can identify the properties of a function, such as its weight, from the expansion. This would considerably narrow down the candidate modular forms or even pinpoint the right modular form that would fit the counting problem under consideration. In what follows, we perform a series of experiments involving known functions of the types encountered in BPS black hole counting problems in superstring theories. We employ a fully connected feed-forward neural network (NN) to determine the weight from the Fourier coefficients. In Appendix~\ref{app:af}, we review the definitions of the automorphic forms that we study. In Appendix~\ref{app:ml}, we specify the machine learning architecture used in our investigations.

\subsection*{Dedekind eta function} 

\begin{table}
    \centering
    \ra{1.5}
    \begin{tabular}{@{}c@{}ccccc@{}}
    \toprule
    \toprule
    \multirow{2}{*}{Experiments} & Data & \multirow{2}{*}{$(w, n_{\text{coeff}})$} & Training & Val.\ & Test  \\[-.75em]
     & size & & Loss & Loss & Error(\%)     \\
    \midrule
    \midrule
    Half-integer negative powers & 400 & $(-200,30)$ & $3.50\times10^{-2}$ & $2.92\times10^{-2}$ & $0.16\%$         \\  \hline
    Random real negative powers & 500 & $(-300,50)$ & $1.45\times10^{-2}$ & $1.93\times10^{-2}$ & $0.21\%$         \\  \hline
    Half-integer positive powers & 400 & $(200,30)$ & $5.18\times10^{-2}$ & $5.05$ & $35.59\%$         \\  \hline
    Half-integer negative powers & \multirow{2}{1.5em}{400} & \multirow{2}{4.5em}{$(-200,20)$} & \multirow{2}{*}{$3.3\times10^{-3}$} & \multirow{2}{*}{$7.94\times10^{-3}$} & \multirow{2}{2em}{$0.04\%$}         \\[-.75em]  
    with random slice of coefficients & & & & &         \\ 
    \bottomrule
    \bottomrule
    \end{tabular}
    \capt{Results from training neural networks using Fourier coefficients obtained from $\eta^n$. The first column of the table specifies the types of values $n$ was allowed to take in various experiments. The second column gives the number of modular forms that comprised the total sample set. The third column specifies the minimum or maximum weight of the modular forms in the dataset and the number of coefficients we used to train the neural network. The range of weights starts at $\pm\frac12$, depending on the sign of the weights being considered. The rest of the columns provide values of the training, validation and test errors, respectively. For test errors, we present the mean relative errors of predictions.} \label{table:delta}
\end{table}

We generated the Fourier coefficients of multiple rational powers of the Dedekind eta function, $\eta(\tau)$. These data in the form~\eqref{eq:dataRep} were then used as inputs to \texttt{Net1} given in~\eref{eq:net1}, a five-layer neural network that predicts the weight.
The results are summarized in Table~\ref{table:delta}. The test errors were quite low, and we conclude that neural networks learn to predict the weights of modular forms of this class meaningfully.

Note that in all of our experiments, the modular forms in our dataset have only a finite range of weights. We have checked that the performance of the trained neural networks is poor on modular forms with weights outside of these ranges. This is expected from the universal approximation theorem, which states that a single-layer neural network of finite width can approximate to arbitrary precision a (suitably well-behaved) function on a compact subset of $\mathbb{R}^n$. The generalization properties of a neural network outside the domain used for training are typically poor.

\subsection*{Mock-modular forms}  

\begin{table}
    \centering
    \ra{1.5}
    \begin{tabular}{@{}c@{}cccc@{}}
    \toprule
    \toprule
    {Experiments} & {$(w_{\text{min}},w_{\text{max}},n_{\text{coeff}})$} & {Training Loss} & {Val.\ Loss} & {Test Error(\%)} \\
    \midrule
    \midrule
    Negative 30 coefficients & $(-200,-\frac12,30)$ & $5.63\times10^{-4}$ & $4.31\times10^{-3}$ & 0.04 \\
    Starting at 11th coefficient & $(-200,-\frac12,30)$ & $9.86\times10^{-3}$ & $2.36\times10^{-2}$ & 0.14 \\
    \hline
    Positive 30 coefficients & $(\frac12,100,30)$ & $2.91\times10^{-1}$ & $7.44\times10^{2}$ & $383.27$ \\
    Calc. Kloosterman sums & $(-100,-\frac12,30)$ &  $9.83\times10^{-4}$ & $9.05\times10^{-4}$ & 0.03 \\
    \hline
    Absolute value of $I_{-w-1}$ & $(-100,-\frac12,33)$ & $1.16\times10^{-3}$ & $1.6\times10^{-3}$ & 0.01 \\
    Real value of $I_{-w-1}$ & $(-100,-\frac12,33)$ & $1.23\times10^{-3}$ & $1.77\times10^{-1}$ & 0.09 \\
    Absolute value of $I_{-w}$ & $(-100,-\frac12,33)$ & $1.1\times10^{-3}$ & $1.57\times10^{-3}$ & 0.04 \\
    Real value of $I_{-w}$ & $(-100,-\frac12,33)$ & $1.97\times10^{-3}$ & $2.07\times10^{-1}$ & 0.17 \\
    \bottomrule
    \bottomrule
    \end{tabular}
    \capt{Results from training neural networks on powers of $E_2$. As in the previous table, the first column specifies the choice of data used to train neural networks. In the second column, $w_{\text{min/max}}$ denotes the minimum/maximum weights of the modular forms in the dataset, and $n_{\text{coeff}}$ denotes the number of coefficients of the Fourier series used. The weights were changed in each case by $\pm\frac12$, so the data size in each experiment is determined from $w_{\text{min/max}}$. The rest of the columns are as in the previous table.} 
    \label{table:mmf}
\end{table}

For the case of mock-modular forms, we study quasi-modular forms constructed from the Eisenstein function $E_2(\tau)$ of weight $2$ and depth $1$. We generate the data similarly to the modular form case.
We explicitly calculated the Fourier coefficients from $2 E_2(q)\Delta(q)^{w/12}$ 
by considering different (both positive and negative) rational values for $w$. The factor of two ensures proper normalization. We also calculated the Fourier coefficients via Kloosterman sum formul\ae, and trained on these data.

We again used \texttt{Net1} to train on this dataset. We confirmed that neural networks also learn to predict weights in this case. We then tried to check if the neural network could predict weights from just the individual parts of~\eqref{eq:Rademacher}, \textit{i.e.}, from the term with either $I_{-w-1}$ or $I_{-w}$. We find that given only absolute values or real parts of the Fourier coefficients of the individual terms, neural networks manage to predict weights well. Our results for this set of experiments are reported in  Table~\ref{table:mmf}. We found that performance was poor when given imaginary parts of the Fourier coefficients and do not report on these experiments.

\subsection*{Jacobi forms}

We consider the most generic Jacobi form, \textit{viz.}, $\theta_1^k \theta_2^l \theta_3^m \theta_4^n$, with $ k, l, m, n  \in\mathbb{Z}$, and use them to generate modular forms by expanding in a Fourier series in $u=e^{2\pi iz}$. The Jacobi theta functions are the elliptic analogues of the exponential function. They satisfy myriad identities. The coefficient of $u^l$ in $\theta_a^k$ is a modular form of weight $k+l/2$ and can itself be expanded in a series in powers of $q=e^{2\pi i \tau}$. We represent these modular forms as data for neural networks as in~\eqref{eq:dataRep}. For $\theta_{3,4}$, the modular form corresponding to the $u^0$ term is discarded for convenience with data manipulation. As separate experiments, we generated and used the modular forms from the functions $\theta_1^k \theta_2^l \theta_3^m \theta_4^n/\eta^j$, for appropriate powers of $\eta$. Since ${\theta_1}$ is similar to ${\theta_2}$ in nature,\footnote{As seen from~\eqref{eq:defn-q1} to~\eqref{eq:defn-q4}, $\theta_{1,2}$ have an extra factor of $q^{{1 \over 8}} e^{\pm i\pi z}$ compared to $\theta_{3,4}$.} and ${\theta_3}$ is similar to ${\theta_4}$, we further consider their individual products with different powers, \textit{i.e.}, $\theta_1^k \theta_2^l$ and $\theta_3^m \theta_4^n$. Further, we trained on the modular forms generated from powers of individual $\theta_a$ with and without division by $\eta$. For this case, instead of using $L^2$-normalization, we use the {encoder}~\eqref{eq:encoder}. We also find \texttt{Net2} in~\eref{eq:net2} to be generally better suited to training on modular forms generated from $\theta_a$. Our results are presented in Tables~\ref{table:theta_individual},~\ref{table:theta_individual_by_delta},~\ref{table:theta_products},~\ref{table:theta_sum}.

\begin{table}
    \centering
    \ra{1.5}
    \begin{tabular}{@{}cccccc@{}}
    \toprule
    \toprule
   {Expts.} & Data size\!\!\! & {$(k_{\text{min}},k_{\text{max}},n_q,n_{u},n_{\text{c}})$} & {Training Loss} & {Val.\ Loss} & {Test Error(\%)} \\
    \midrule
    \midrule
    & 1360 & $(1,50,70,80,26)$ & $3.60\times10^{-2}$ & $4.83\times10^{-2}$ & $0.26$ \\
    $\theta_1^k$ & 1640 &  $(-40, -1,50,60,26)$ & $3.90\times10^{-2}$ & $5.15\times10^{-1}$ & $4.54$ \\
    & 1808 & $(-40,40,50,60,25)$ & $9.71\times10^{-3}$ & $5.49\times10^{-1}$ & $6.13$ \\
    \hline
    & 2009 & $(1,50,70,80,26)$ & $5.73\times10^{-2}$ & $2.12\times10^{-2}$ & $0.12$ \\
    $\theta_2^k$ & 1240 & $(-40, -1,50,60,26)$ & $1.59\times10^{-1}$ & $8.74\times10^{-1}$ & $2.48$ \\
    & 2170 & $(-40,40,50,60,22)$ & $2.91\times10^{-2}$ & $6.74\times10^{-2}$ & $2.16$ \\
    \hline
    & 1170 & $(1,40,50,60,24)$ & $9.53\times10^{-4}$  & $5.85\times10^{-4}$ & $0.13$ \\
    $\theta_3^k$ & 1200 & $(-40, -1,50,60,50)$ & $1.68\times10^{-2}$ & $3.91\times10^{-2}$ & $0.54$ \\
    & 2370 & $(-40,40,50,60,24)$ & $2.42\times10^{-2}$ & $7.48\times10^{-2}$ & $0.31$ \\
    \hline
    & 1170 & $(1,40,50,60,24)$ & $7.47\times10^{-4}$ & $7.55\times10^{-4}$ & $0.10$ \\
    $\theta_4^k$ & 1200 & $(-40, -1,50,60,50)$ & $8.47\times10^{-2}$ & $3.46\times10^{-2}$ & $0.80$ \\
    & 2340 & $(-40,40,50,60,24)$ & $4.99\times10^{-3}$ & $1.83\times10^{-2}$ & $0.38$ \\
    \bottomrule
    \bottomrule
    \end{tabular}
    \capt{Results for modular forms generated by taking various powers of the individual Jacobi theta functions. $k_{\text{min/max}}$ denote the minimum and maximum powers of $\theta_i$ used. We consider Fourier expansions up to maximum powers $n_{q,u}$ of $q,u$, using only modular forms that have at least $n_{\text{c}}$ non-zero coefficients in their expansions.}
    \label{table:theta_individual}
\end{table}

In Table~\ref{table:theta_individual}, we use powers of the individual $\theta_a$ to generate modular forms to train neural networks.
We expand each $\theta_a^k$ up to some powers $n_u$ of $u$ and $n_q$ of $q$, to generate a collection of modular forms as described just above.
Note that $k$ ranges between specified values of $k_{\text{min}}$ and $k_{\text{max}}$. From the generated modular forms, we select those that have at least some chosen number $n_c$ of non-zero coefficients in their Fourier expansions up to $\mathcal{O}(n^q)$. This is the reason why the data size varies between experiments. We trained neural networks on datasets generated using just positive powers of $\theta_a$, using just negative powers of $\theta_a$, and using both positive and negative powers of $\theta_a$. We find that in experiments with $\theta_{1,2}$, there is a significant difference in the performance of neural networks between the positive and negative powers. No such difference is seen when training on modular forms generated using positive or negative powers of $\theta_{3,4}$.

\begin{table}
    \centering
    \ra{1.5}
    \begin{tabular}{@{}cccccc@{}}
    \toprule
    \toprule
    {Expts.} & Data size\!\!\! & {$(k_{\text{min}},k_{\text{max}},n_q,n_{u},n_{\text{c}})$} & {Training Loss} & {Val.\ Loss} & {Test Error(\%)} \\
    \midrule
    \midrule
    & 1350 & $(1,50,70,80,26)$ & $2.63\times10^{-2}$ & $3.54\times10^{-2}$ & $0.38$ \\
    $\theta_1^k/\eta^{3k}$ & 1600 & $(-40, -1,50,60,25)$ & $3.35\times10^{-2}$ & $1.36\times10^{-2}$ & $0.17$ \\
    & 2380 & $(-40,40,50,60,25)$ & $7.73\times10^{-3}$ & $1.22\times10^{-2}$ & $0.32$ \\
    \hline
    & 2050 & $(1,50,70,80,26)$ & $7.18\times10^{-2}$ & $6.39\times10^{-2}$ & $0.89$ \\
    $\theta_2^k/\eta^{3k}$ & 2050 & $(-50, -1,70,80,35)$ & $1.97\times10^{-1}$ & $8.45\times10^{-1}$ & $0.46$ \\
    & 4098 & $(-50,50,70,80,35)$ & $1.69\times10^{-1}$ & $7.81\times10^{-1}$ & $1.06$ \\
    \hline
    & 1200 & $(1,40,50,60,24)$ & $4.93\times10^{-3}$  & $5.79\times10^{-3}$ & $0.29$ \\
    $\theta_3^k/\eta^{12}$ & 1200 & $(-40, -1,50,60,51)$ & $2.39\times10^{-3}$ & $1.18\times10^{-2}$ & $0.24$ \\
    & 2400 & $(-40,40,50,60,50)$ & $2.18\times10^{-1}$ & $1.4\times10^{-1}$ & $0.41$ \\
     \hline
    & 1200 & $(1,40,50,60,24)$ & $2.41\times10^{-3}$ & $3.16\times10^{-3}$ & $0.49$ \\
    $\theta_4^k/\eta^{12}$ & 1200 & $(-40, -1,50,60,51)$ & $1.72\times10^{-2}$ & $1.75\times10^{-2}$ & $0.44$ \\
    & 2400 & $(-40,40,50,60,50)$ & $5.56\times10^{-2}$ & $1.13\times10^{-1}$ & $0.48$ \\
    \bottomrule
    \bottomrule
    \end{tabular}
    \capt{Results for modular forms generated by taking various powers of the individual Jacobi theta functions divided by appropriate powers of $\eta$ to ensure that the Fourier expansion starts at $q^0$, except for the bottom two sub-rows of the row $\theta_2^k/\eta^{3k}$, where we started at $q^1$. $k_{\text{min/max}}$ denote the minimum and maximum powers of $\theta_i$. We consider Fourier expansions up to maximum powers $n_{q,u}$ of $q,u$. We only use modular forms that have at least $n_{\text{c}}$ non-zero coefficients in their expansions.}
    \label{table:theta_individual_by_delta}
\end{table}

Table~\ref{table:theta_individual_by_delta} presents results for training on modular forms generated from powers $(\theta_{1,2}/\eta^3)^k$ and $\theta_{3,4}^k/\eta^{12}$, with $k$ again ranging between specified values $k_{\text{min}}$ and $k_{\text{max}}$.
The different powers of $\eta$ used for division are chosen to ensure that the Fourier series in each case starts at $q^0$. For this set of experiments, we find good performance of the trained neural networks for each experiment, independent of the function $\theta_a$ and the signs of their powers.

\begin{table}
    \centering
    \ra{1.5}
    \begin{tabular}{@{}lrcccc@{}}
    \toprule
    \toprule
    {Expts.} &\ \  Data size\!\!\!\! & {$(k_{\text{min}},k_{\text{max}},n_q,n_{u},n_{\text{c}})$} & {Training Loss} & {Val.\ Loss} & {Test Error(\%)} \\
    \midrule
    \midrule
    \multirow{2}{2pt}{$\theta_1^k\theta_2^l$} & 14193 & $(1,20,70,80,26)$ & $3.14\times10^{-2}$ & $5.12\times10^{-2}$ & $0.35$ \\
    & 803 & $(-5, -1,50,60,26)$ & $6.67\times10^{-1}$ & $9.68\times10^{-1}$ & $7.05$ \\
    \hline
    \multirow{2}{2pt}{$\theta_3^k\theta_4^l$} & 15960 & $(1,20,70,80,35)$ & $2.91\times10^{-3}$ & $2.4\times10^{-3}$ & $0.08$ \\
    & 750 & $(-5, -1,50,60,25)$ & $3.88\times10^{-2}$ & $1.4\times10^{-2}$ & $0.67$ \\
    \midrule
    \multirow{2}{0pt}{$\theta_1^k\theta_2^l\theta_3^m\theta_4^n$} & 19598 & $(1,5,70,80,53)$ & $7.88\times10^{-3}$ & $3.31\times10^{-3}$ & $0.11$ \\
    & 1416 & $(-2, 0,50,60,40)$ & $2.78\times10^{-2}$ & $8.40\times10^{-2}$ & $1.15$ \\
    \hline
    \multirow{2}{2pt}{$\displaystyle\frac{\theta_1^k\theta_2^l\theta_3^m\theta_4^n}{\eta^j}$} & 18239 & $(1,5,50,60,25)$ & $1.54\times10^{-1}$ & $6.4\times10^{-1}$ & $2.79$ \\
    & 1416 & $(-2, 0,40,50,39)$ & $2.81\times10^{-2}$ & $8.38\times10^{-2}$ & $1.15$ \\
    \bottomrule
    \bottomrule
    \end{tabular}
    \capt{Results for modular forms generated by taking various powers of products of Jacobi theta functions in specified combinations. $k_{\text{min/max}}$ denote the minimum and maximum powers of $\theta_i$ used to generate the modular forms. We consider Fourier expansions up to maximum powers $n_{q,u}$ of $q,u$. We only use modular forms that have at least $n_{\text{c}}$ non-zero coefficients in their expansions.}
    \label{table:theta_products}
\end{table}

In Table~\ref{table:theta_products}, we present results for experiments with products of powers of different $\theta_a$ without and with division by $\eta$. Note that in each data set, we used significantly more modular forms for the case with positive powers of $\theta_a$. This is due to the fact that generating Fourier series for negative powers of $\theta_a$ is considerably slower in \textsf{Mathematica} compared to positive powers. The first two sets of experiments are with the products $\theta_1^k \theta_2^l$ and $\theta_3^k \theta_4^l$. For experiments with modular forms generated from the products $\theta_1^k \theta_2^l$, we find low errors when considering only positive powers, while the performance has high errors for the data set with only negative powers. For the products $\theta_3^k \theta_4^l$, errors are quite low for both positive and negative powers. The lower errors in experiments with $\theta_3^k \theta_4^l$ compared to $\theta_1^k \theta_2^l$ are probably due to the extra oscillating factor of $q^{\frac{1}{8}}\,e^{\pm i\pi z}$ in $\theta_{1,2}$. For experiments with the products $\theta_1^k\theta_2^l\theta_3^m\theta_4^n$, we find good performance when using only positive powers and slightly worse results when using only negative powers. We also use the combination $\theta_1^k\theta_2^l\theta_3^m\theta_4^n/\eta^j$ to train neural networks. Here, $j=\frac{1}{2} \left(\frac{k+l}{4}+\alpha-\beta\right)$, with $\alpha=1$ if $m$ or $n$ are non-zero and $\alpha=0$ otherwise, and $\beta$ is the desired power of the first term $q^\beta$ in the Fourier expansion. We find that this combination leads to somewhat reduced performance compared to experiments without division by $\eta^j$.

\begin{table}
    \centering
    \ra{1.5}
    \begin{tabular}{@{}lrcccccc@{}}
    \toprule
    \toprule
    {Expts.} & Data size\!\!\! & {$(k_{\text{min}},k_{\text{max}},n_q,n_{u},n_{\text{c}}, \sigma)$} & {Training Loss} & {Val.\ Loss} & {Test Error(\%)} \\
    \midrule
    \midrule
    \multirow{2}{1pt}{$\theta_1^k\theta_2^l\theta_3^m\theta_4^n$} & 2867 & $(-5,5,50,70,43,+)$ & $4.66\times10^{-2}$ & $6.2\times10^{-1}$ & $3.89$ \\
    & 7221 & $(-5, 5,50,70,36,-)$ & $3.86\times10^{-1}$ & $7.6\times10^{-1}$ & $6.32$ \\   \midrule
    \multirow{2}{1pt}{$\displaystyle\frac{\theta_1^k\theta_2^l\theta_3^m\theta_4^n}{\eta^{j}}$} & 2867 & $(-5,5,50,70,43,+)$ & $1.87\times10^{-2}$ & $5.57\times10^{-1}$ & $4.22$ \\
    & 978 & $(-5,5,50,70,25,-)$ & $3.63\times10^{-2}$ & $5.28\times10^{-1}$ & $13.34$ \\
    \bottomrule
    \bottomrule
    \end{tabular}
    \capt{Results for modular forms generated by taking various powers of products of Jacobi theta functions in specified combinations given a choice of $\sigma=\text{sgn}(k+l+m+n)$. $k_{\text{min/max}}$ denote the minimum and maximum powers of $\theta_i$ used to generate the modular forms. We consider Fourier expansions up to maximum powers $n_{q,u}$ of $q,u$. We only use modular forms that have at least $n_{\text{c}}$ non-zero coefficients in their expansions.}
    \label{table:theta_sum}
\end{table}

Finally, Table~\ref{table:theta_sum} contains results for experiments where modular forms were generated by considering both positive and negative powers in $\theta_1^k\theta_2^l\theta_3^m\theta_4^n$ while ensuring that $s=k+l+m+n$ has a fixed sign, $\sigma=\text{sgn}(s)$, in each data set. We further considered the combination $\theta_1^k\theta_2^l\theta_3^m\theta_4^n/\eta^j$ subject to the same constraint on $\sigma$, and $j$ as defined previously. The results for these experiments are consistently worse than other experiments.

\section{Summary and prospects} \label{sec:summary}

In this work, we have trained artificial neural networks to predict the weights of modular forms and mock-modular forms given some of the first few Fourier coefficients of the (mock) modular forms. We generated the modular forms used to train and test the networks using various automorphic forms, including the Dedekind eta function, the Eisenstein series $E_2$, and the four elliptic Jacobi theta functions. The results are presented in Tables~\ref{table:delta}-\ref{table:theta_sum}. Note that, since this work is primarily intended as a proof of concept, we have worked with datasets that are not too large in size. 

We found that neural networks learn to predict the weights of negative powers of Dedekind eta (with negative weights) quite well.
The same is true for negative powers of $E_2$, which again have negative weights. In both these cases, the performance degrades severely for positive weight (mock-)modular forms. We generated modular forms using Jacobi theta functions in multiple ways, detailed in Section~\ref{sec:results} and Tables~\ref{table:theta_individual}-\ref{table:theta_sum}. With these modular forms, we find in most of the experiments with just $\theta_{1,2}$ that the performance is better for positive powers and hence positive weight modular forms. For experiments with powers of $\theta_{3,4}$, we find no significant difference between positive and negative weights. This difference between $\theta_{1,2}$ and $\theta_{3,4}$ might be attributable to the absence of negative powers of $u$ in $\theta_{3,4}$. When we use the products of the powers of the four $\theta_a$ together, the predictions of the neural networks are not as good compared to the other cases. It would seem that the presence of multiple distinct $\theta_a$ together makes learning for neural networks more difficult.

In the case of superstring theory, modular, Jacobi and quasi-modular forms appear prominently as counting functions for BPS black holes.
The machine learning approach developed here has the potential to determine the weight of modular forms in cases where they are only incompletely known (\textit{e.g.}, only a finite number of Fourier coefficients have been identified). As an outstanding example, in the case of the $\mathcal{N}=2$ four-dimensional superstring STU model, the known approximate counting function for $\frac12$-BPS black holes~\cite{Cardoso:2019avb} with a non-zero classical horizon area suggests that the exact counting formula is a combination of a Siegel modular form as well as an infinite convergent series of functions. The former has been precisely identified, while the latter is suggestive of a Jacobi form expanded in the elliptic variable. The machine learning approach here is well-suited to determining the weight of this putative Jacobi form, thereby narrowing the space of candidate Jacobi forms containing the right one for the counting function. An immediate future direction would be to apply our protocol to congruent subgroups of $SL(2,\mathbb{Z})$ which constitute the symmetry groups governing various compactification models of interest for black hole counting, such as the CHL models~\cite{Jatkar:2005bh} in superstring theory, so as to be able to probe for automorphic forms in these models. We will report on this in forthcoming work.

Holographic dualities such as AdS/CFT offer an ingenious pathway to represent gravitational states and interactions in a quantum field theoretic framework. This is notable, not least because the gauge/gravity correspondence gives a schema for representing a gravitational system in terms of codimension one data, and this is, in principle, the most efficient encoding available. Proving the duality necessarily requires the exact matching of rigorous calculations on both sides of the correspondence. BPS black hole counting functions, which have been exactly written down in various cases, offer a ready laboratory within which to compare exact results in gravity and CFT. In this setting, the matching of the counting problem reduces to the identification of automorphic forms on either side of the duality. Starting with~\cite{Hashimoto:2018ftp, Hashimoto:2019bih}, there have been suggestions for using deep learning to encode the AdS/CFT correspondence. In the black hole context, aspects of the dictionary have been examined by learning features of bulk horizons from boundary data~\cite{Jejjala:2023zxw}. Machine learning driven approaches, such as the one in this paper, provide a pathway to a new perspective on holography and have the potential to make a decisive contribution to generating exact results in AdS/CFT.

\section*{Acknowledgements}
We thank Gabriel Lopes Cardoso, Yang-Hui He, Dami\'an Mayorga Pe\~na, and Challenger Mishra for discussions. PR would like to thank Chennai Mathematical Institute for hospitality while this work was being carried out. VJ and DN are supported by the South African Research Chairs Initiative of the Department of Science, Technology, and Innovation and the National Research Foundation (NRF), grant 78554. PR is supported by an NRF Freestanding Postdoctoral Fellowship. AS is supported by NRF through the grant PMDS22061422341.

\appendix

\section{Automorphic forms}\label{app:af}

Here, we collect definitions of all the functions that we use.

\subsection*{Modular forms}

The Dedekind eta function is defined, with $q=e^{2\pi i\tau}$, to be
\begin{equation}        \label{eq:dedkind-eta}
    \eta(\tau) = q^{1/24} \prod_{n\geq 1} \left( 1-q^n \right)~.
\end{equation}
$\eta$ is a modular form of weight $\frac12$, and its powers $\eta^k$ are modular forms with weight $\frac{k}{2}$ for $k\in\mathbb{R}$. 
The discriminant function, $\Delta(\tau)$, is
\begin{equation}
\Delta(\tau) = \eta(\tau)^{24} = q \prod_{n=1}^{\infty} \left( 1-q^n \right)^{24}~.
\end{equation}
A holomorphic modular form, $h(\tau)$, of negative weight $w$ has the Fourier expansion
\begin{equation}
    h(\tau) = \sum_{n=0}^\infty \Omega(n) q^{n-\lambda}~,
\end{equation}
with $\lambda=-\frac{1}{12} w$, and has the polar part
\begin{equation}
    h^-(\tau) = \sum_{n-\lambda<0}^\infty \Omega(n) q^{n-\lambda}~.
\end{equation}
The Fourier coefficients with $n\geq\lambda$ of $h(\tau)$ can be written in terms of the polar coefficients using a Rademacher expansion,
\begin{equation}        \label{eq:kloosterman}
    \Omega^{(w)}(n) = 2 \pi \sum_{n' < \lambda} \Omega(n') \sum_{c=1}^{\infty} \frac{1}{c} K(n-\lambda, n'-\lambda) \left(\frac{n'-\lambda}{n-\lambda}\right)^{\frac{1-w}{2}} I_{1-w}\left( \frac{4 \pi}{c} \sqrt{(n-\lambda) \abs{n'-\lambda}} \right)~.
\end{equation}
Here, $K(x,y)$, called a Kloosterman sum, is defined by 
\begin{equation}
    K(x,y) = i^{-w} \sum_{\begin{matrix}-c < d < 0 \\  (c,d)=1\end{matrix}} \mathbf{E}\left(\frac{a}{c}y + \frac{d}{c}x\right)~,
\end{equation}
with $ ad=1 \text{mod}\,{c}$, $\mathbf{E}(x) = e^{2\pi i x}$, and $I_\nu(x)$ is the modified Bessel function of the first kind,
\begin{equation}
    I_{\nu}(z) = \left( \frac{z}{2} \right)^{\nu} \sum_{k=0}^{\infty} \frac{z^{2k}}{2^{2k} k! \Gamma(\nu+k+1)}.
\end{equation}

\subsection*{Mock-modular form Eisenstein $E_2$}

The Eisenstein series $E_2$ is defined as follows:
\begin{equation}
    E_2(\tau) = \frac{1}{2\pi i } \frac{\Delta'(\tau)}{\Delta(\tau)} = 1 - 24 \sum_{n=1}^{\infty} \frac{n q^n}{1-q^n}~.
\end{equation}
It is a mock-modular form of weight $2$.
Writing
\begin{equation}
    2 \frac{E_2(\tau)}{\eta^{24}(\tau)} = \sum_{n=-1}^{\infty} c(n) q^n~,
\end{equation}
and using
\begin{equation}
    \frac{E_2(\tau)}{\eta^{24}(\tau)} = - q \frac{d}{dq} \frac{1}{\eta^{24}(\tau)}~,
\end{equation}
the Fourier coefficients $c(n)$ can be written in terms of the Fourier coefficients of $\eta^{-24}(\tau)$ as
\begin{equation}
    c(n) = - 2 n\, d(n)~,
\end{equation}
where $d(n)$ are the Fourier coefficients of $\eta^{-24}(\tau)$.
This can be generalized to determine Fourier coefficients of $E_2(\tau)\eta^{2w}$ for arbitrary negative weights $w$.
Writing
\begin{equation}
    E_2(\tau)\eta^{2w} = \sum_{n\geq-\lambda}^\infty c^{(w)}(n)q^n~, 
\end{equation}
where $\lambda=-\frac1{12}w$, one can obtain the following:
\begin{equation}
    c^{(w)}(n) = \frac{24}{w} n\, d^{(w)}(n)~,
\end{equation}
where $d^{(w)}(n) = \Omega^{(w)}(n+\lambda)$ are the Fourier coefficients~\eqref{eq:kloosterman} of $\eta(\tau)^{-w/12}$.
We also have a Rademacher expansion of $d^{(w)}(n)$,
\begin{equation}
    d^{(w)}(n) = 2 \pi \sum_{n' < \lambda} \Omega(n') \sum_{c=1}^{\infty} \frac{1}{c} K(n, n'-\lambda) \left(\frac{n'-\lambda}{n}\right)^{\frac{1-w}{2}} I_{1-w}\left( \alpha \right)~,
\end{equation}
with $\alpha=\frac{4 \pi}{c}\sqrt{n \abs{n'-\lambda}}$.
Using the recurrence relation of modified Bessel functions, 
\begin{equation}
    I_{m-1}(x) - I_{m+1}(x) = \frac{2m}{x}I_m(x)~,
\end{equation}
we split the Rademacher expansion as
\begin{align}       \label{eq:Rademacher}
    c^{(w)}(n) =&\ \frac{48 \pi}w \sum_{n' < \lambda} \Omega(n') K(n, n'-\lambda) (n'-\lambda)^{\frac{1-w}{2}}    \sum_{c=1}^{\infty} \frac{1}{c}  \left( \frac{I_{-w-1}(\alpha)}{n^{-(w+1)/2}}  + \frac{2w}{\beta}\frac{I_{-w}(\alpha)}{n^{-w/2}} \right)~,
\end{align}
where $\beta = \alpha/\sqrt n$.

\subsection*{Jacobi forms} 

We also use the elliptic Jacobi theta functions, $\theta_a$, $a=1,2,3,4$, to generate modular forms.
They are given, with $q = e^{2\pi i \tau}$, as
\begin{align}
    \theta_1(z,\tau) &= \sum_{n=-\infty}^\infty (-1)^{n-1/2} q^{(n+1/2)^2/2} e^{(2n+1)\pi iz} \, ,  \label{eq:defn-q1} \\
    \theta_2(z,\tau) &= \sum_{n=-\infty}^\infty q^{(n+1/2)^2/2} e^{(2n+1)\pi iz} \, , \label{eq:defn-q2}\\
    \theta_3(z,\tau) &= \sum_{n=-\infty}^\infty q^{n^2/2} e^{2n\pi iz} \, , \label{eq:defn-q3}\\
    \theta_4(z,\tau) &= \sum_{n=-\infty}^\infty (-1)^n q^{n^2/2} e^{2n\pi iz} \, \label{eq:defn-q4}.
\end{align}

These are all Jacobi forms of weight $1/2$ and index $1$, with the first two being distinct from the last two in that they are cuspidal forms that go to zero at $\tau = i\infty$ and have extra oscillatory factors of the form $e^{\pm i \pi z}$.

\section{Data and machine learning}\label{app:ml}

In all of our experiments, the input to neural networks (NNs) for each modular form is schematically
\begin{equation}        \label{eq:dataRep}
    \texttt{\{FourierCoefficients} \to \texttt{Weight\}}
\end{equation}
in \textsf{Mathematica} syntax, where \texttt{FourierCoefficients} are (some of the) first few Fourier coefficients of a (mock) modular form, and \texttt{Weight} is the weight of the corresponding modular form.
The choice of the Fourier coefficients changes between experiments.
The vector of Fourier coefficients that we choose undergoes further processing. When training neural networks on powers of $\eta$ and $E_2$, the vectors of coefficients are normalized using an $L^2$-norm.
In some experiments on modular forms obtained from Jacobi forms, the coefficients are passed through an \texttt{encoder}, an element-wise layer, given as follows in \textsf{Mathematica} syntax as
\begin{equation}        \label{eq:encoder}
    \texttt{encoder} = \texttt{Log[N[Abs[}\bm{\cdot}\texttt{]]]}~,
\end{equation}
where $\bm{\cdot}$ denotes the coefficients.

We use simple feed-forward neural network to perform our machine learning experiments. We use the following networks,
\begin{align}
\texttt{Net1} =&\ (\text{Input}, 256, \text{ReLU}, 128, \sigma, 64, \text{ReLU}, 32, \text{ReLU}, 8, \text{ReLU}, 4, \text{ReLU}, \text{Output})~, \label{eq:net1} \\
\texttt{Net2} =&\ (\text{Input}, 256, \text{ReLU}, 256, \text{GELU}, 128, \text{ReLU}, 128, \text{GELU}, 64, \text{ReLU}, \label{eq:net2} \\
&\ \quad 32, \text{GELU}, 8, \text{ReLU}, 4, \text{ReLU}, \text{Output})~,       \nonumber
\end{align}
where the numbers are the numbers of neurons in each of the hidden layers of the feed-forward neural network, Rectified Linear Unit (ReLU) is the elementwise $\text{max}(0,x)$ non-linearity function, $\sigma(x)=(1+e^{-x})^{-1}$ is the logistic sigmoid function, and GELU is the Gaussian Error Linear Unit, acting as $\frac x2(1+\text{erf}(\frac x{\sqrt 2}))$. We use \texttt{Net1}, especially for $\eta$ and $E_2$.
We observe that \texttt{Net2} performs comparatively better than \texttt{Net1} for our purposes. This is likely due to the higher number of layers and the fact that GELU performs better when dealing with data of different orders of magnitude. As this paper constitutes a proof of principle of the utility of neural network machine learning for the counting functions relevant to black holes, we have not optimized the architecture.

For purposes of reproducibility, let us elaborate on the training protocol. In our machine learning experiments, we break our training data in mini-batches of length $64$. We use the Adaptive Moment Estimation (ADAM) algorithm with a learning rate $\eta = 0.001$, with the mean squared error (MSE) as the loss function. Mean relative errors are used to compare predicted outputs with actual outputs. Although the number of modular forms used varies between experiments, we always use a $$(75\%,15\%,10\%)$$ split to form the training, validation, and test sets, respectively.

Finally, we note that our datasets for experiments on modular forms generated from $\theta_a$ are smaller when considering negative powers despite poor results. This is because generating the Fourier series in these cases is significantly more time-consuming than in the other cases.

Our datasets and code are available on \href{https://github.com/abinash7s/ml_modular_forms}{\textsf{Github}}~\cite{abinash7smlmf}.

\providecommand{\href}[2]{#2}\begingroup\raggedright\endgroup

\end{document}